\documentclass[10pt, conference, compsocconf]{IEEEtran}
\IEEEoverridecommandlockouts

\usepackage{cite}
\usepackage{amsmath,amssymb,amsfonts}
\usepackage{algorithmic}
\usepackage{graphicx}
\usepackage{textcomp}
\usepackage{xcolor}
\def\BibTeX{{\rm B\kern-.05em{\sc i\kern-.025em b}\kern-.08em
    T\kern-.1667em\lower.7ex\hbox{E}\kern-.125emX}}
\begin{document}

\title{A Cross-Cultural Analysis of Animated Representations of Emotions for Wearable Interfaces\\
\thanks{This work was partially financially supported by Gdańsk University of Technology under the grant DEC-35/1/2024/IDUB/IV.2a/Eu within the Europium -- ‘Excellence Initiative - Research University’ program.}
}

\author{\IEEEauthorblockN{Michal R. Wrobel}
\IEEEauthorblockA{\textit{Faculty of Electronics,}\\
\textit{Telecommunications and Informatics,} \\
\textit{and Digital Technologies Center} \\
\textit{Gdansk University of Technology, }\\
Gdansk, Poland \\
michal.wrobel@pg.edu.pl}
\and
\IEEEauthorblockN{Duygun Erol Barkana}
\IEEEauthorblockA{\textit{Faculty of Engineering, Department of} \\
\textit{Electrical and Electronics Engineering} \\
\textit{Yeditepe University} \\
Istanbul, Turkey \\
duygunerol@yeditepe.edu.tr}
\and
\IEEEauthorblockN{Agnieszka Landowska}
\IEEEauthorblockA{\textit{Faculty of Electronics,}\\
\textit{Telecommunications and Informatics} \\
\textit{Gdansk University of Technology, }\\
Gdansk, Poland \\
nailie@pg.edu.pl}
}

\maketitle
\textbf{ Preprint Notice}
This manuscript is a preprint of an article accepted for presentation at the \textit{EmotionSense – 1st International Workshop on Human-centred Emotion-Awareness and Sensing in Pervasive Computing within The 24th International Conference on Pervasive Computing and Communications (PerCom 2026)}, Pisa, Italy. The final published version may differ as a result of the publisher’s production process.\\[2mm]

\begin{abstract}
Although pervasive sensing technologies are increasingly capable of continuously detecting human emotional states, there is still a critical challenge: how to unobtrusively communicate this sensed data back to the user. Realistic avatars are effective but often unsuitable for the limited screen space and peripheral nature of wearable. Abstract geometric animation offers a promising, rapidly interpretable alternative, but its cross-cultural validity remains under-explored. This study investigates the universality of animated emotion representations. We conducted a comparative study with 105 participants from Poland and Turkey and analyzed how they map emotions to visual parameters, such as color, shape, size, speed, and animation type. The results indicate that color and object size are universally understood as carriers of emotional meaning, making them suitable for global visualization models. However, some cultural variation in dynamic range preferences was revealed by animation speed. These results lay the groundwork for developing generative visualization algorithms that translate continuous sensor data into intuitive, culturally relevant feedback for pervasive environments.
\end{abstract}


\section{Introduction}
Thanks to the ubiquity of wearable sensors and affective computing, systems can now detect human emotional states continuously. However, effectively communicating sensed data back to the user remains a critical design challenge for these pervasive systems. Although verbal and facial expressions have been extensively studied, visualizing sensed affect through abstract animated representations remains an under-explored area, especially with regard to its cross-cultural validity in global applications.

Culture plays a significant role in how emotions are expressed, perceived, and interpreted~\cite{Srinivasan2018Dec}. Differences in language, social norms, and artistic traditions influence the way people associate meaning with visual cues like color and motion. As pervasive computing systems are deployed globally, determining whether these cultural differences extend to abstract feedback is essential. If substantial cross-cultural differences exist, visualization algorithms must be systematically adapted to ensure that emotion-based notifications remain effective and interpretable across diverse user groups. Understanding these differences is essential to assessing whether a universal visualization model is feasible or if culturally adaptive algorithms are required~\cite{Barrett2011Oct}.

Various approaches have been developed to visualize emotions, ranging from facial expression synthesis~\cite{Yoshitomi2021Jun} and virtual avatars~\cite{Pan2024Mar} to abstract representations using colors~\cite{Demir2020Oct}, shapes~\cite{Wang2023Feb}, and motion~\cite{Wrobel2024}. Although realistic representations are effective in immersive environments, they are often not suitable for pervasive computing scenarios. Realistic avatars can be intrusive, privacy-invasive, or cognitively demanding on small wearable displays or ambient home devices. In contrast, abstract representations, such as simple animations of geometric shapes, offer an easily digestible alternative. These representations allow systems to unobtrusively convey affect by focusing on basic visual properties such as speed, color, and size~\cite{Wrobel2024}.

The main objective of this study is to investigate whether cultural background influences how individuals perceive these animated indicators of emotion. For this purpose, an empirical study was conducted in Turkey and Poland involving student participants from both countries. Through the analysis of visual parameters such as color, shape, animation type, speed, and size, we aimed to identify universal patterns versus culturally specific preferences.

\section{Related work}\label{sec:rw}
The transition from detecting human emotions using sensors to effectively conveying this information to the user poses a serious challenge in the field of ubiquitous computing. While sensor accuracy has improved, the design of affective feedback interfaces, particularly for mobile and wearable devices, requires visualization techniques that are interpretable yet unobtrusive. Research has explored the role of design features such as color, motion, and interactivity in conveying emotion~\cite{Wang2023Feb}. However, while emotion visualization is a growing field, relatively little attention has been paid to the cross-cultural validity of these feedback mechanisms, especially when using abstract animation.

\subsection{Visualizing Sensed Affect: From Realism to Abstraction}
The visualization of sensed emotions can be broadly divided into two categories: realistic (e.g., avatars) and abstract (e.g., geometric shapes). While facial expression synthesis~\cite{Yoshitomi2021Jun} and virtual avatars~\cite{Pan2024Mar} provide high nuance, they can be cognitively demanding or socially intrusive when displayed on pervasive devices like smartwatches. Sonderegger et al. demonstrated that animated abstract tools (e.g., ``AniSAM'') can convey affective states as accurately as anthropomorphic representations, with superior performance in low-arousal conditions~\cite{Sonderegger2016May}. This suggests that abstract geometric animations may be particularly suitable for ambient displays and wearables, where screen real estate is limited and ``glanceability'' is paramount.

Designing effective abstract feedback requires an understanding of how basic visual variables map to emotion. In the context of static data, Blair et al. and Anderson et al. demonstrated that design elements such as color schemes, data density, and variability can significantly influence how emotion is perceived, even when the data itself has no semantic meaning~\cite{blair2024quantifying, Anderson2022}.

While these studies provide a foundation for static visualization, pervasive sensing requires dynamic feedback. Lan et al. reviewed 109 papers on emotion visualization, noting that while realistic imagery is common, kinetic motion remains an underutilized but powerful modality for evoking emotion~\cite{lan2023affective}. Wrobel et al. began to address this by identifying typical representations of discrete emotions in terms of animation's speed, size, and shape~\cite{Wrobel2024}. However, existing models for animated feedback have not yet been systematically stress-tested across different cultural groups.

\subsection{Cultural Context in Pervasive Design}
For pervasive computing systems to be universally applicable, visualization algorithms must account for the user's cultural background. Differences in language, norms, and artistic traditions influence how people associate meaning with visual cues.

In the visual domain, color-emotion associations have been extensively mapped. Jonauskaite et al. analyzed responses from 30 nations, finding a high degree of universal consistency (e.g., red/black with anger) but notable local variations in the strength of these associations~\cite{Jonauskaite2020Sep}. Similarly, Gao et al. found that while the emotional interpretation of chroma and lightness is largely consistent, cultural groups differ on whether they prioritize hue or chroma when assessing ``warmth'' or ``coolness''~\cite{Gao2007Jun}.

Cultural divergences become more pronounced when moving beyond color to shape and size, which are critical components of abstract animation. Sanchez et al. investigated visualizations for long-distance interaction and found that, although most emotions were consistently mapped to three-dimensional shapes across Asian and Latin American participants, the visualization of sadness varied significantly between cultures~\cite{Sanchez2014}. 

Despite these findings, the intersection of \textit{animation parameters} (speed, motion type) and \textit{cultural background} remains underexplored. To create robust human-centered sensing systems, it is necessary to determine which parameters of animated feedback are universal and which must be adapted to the local context.

\section{Study design}\label{sec:rm}
For the purpose of the study an interactive web-based questionnaire was developed to determine which parameters of animated emotion representation remain consistent across cultures and which show significant cross-cultural variation. 


During the study, participants were asked to create animated representations of emotions by selecting visual properties that best conveyed each emotion. Each participant was given a set of emotion labels and asked to create an animation for each one. They selected a geometric shape from a predefined set that included a circle, rectangle, diamond, triangle, oval, and star. They also selected an animation type from seven options: bounce, rotate, fade, shake, scale, flip, and stretch as well as specified the speed of the animation and the size of the shape.

To define color, participants used an RGB-based color picker that allowed them to select any shade within the RGB spectrum. To allow for a more precise analysis of color-emotion associations, the originally collected RGB color values were converted into the HSV (hue, saturation, value) model. The HSV model provides a more perceptually relevant representation of color, where hue corresponds to the type of color, saturation reflects its intensity, and value determines its brightness. This transformation enabled a more nuanced investigation of how participants associate specific colors with emotions.

\section{Results}\label{sec:results}
A total of 105 participants completed the survey, consisting of 51 students from Poland and 54 students from Turkey. All participants were within a similar age range of 19 to 23 years old, ensuring a relatively homogenous sample in terms of age and educational background. The distribution of participants across the two countries allowed for a comparative analysis of cultural differences in the selection of animation parameters for representing emotions. The following section presents the results of this analysis, focusing on both cross-cultural differences and the relationships between animation parameters and emotion labels.

\subsection{Cultural differences}
In order to determine whether cultural background influences the selection of animation parameters for the representation of emotions, a comparative analysis was conducted between Polish and Turkish participants. The study examined differences in color hue, saturation, and value, as well as shape, animation type, speed, and size. These animation parameters were considered universal if statistical tests did not reveal significant differences between Polish and Turkish participants.

\begin{table}[tb]
\centering
\caption{Independence of color dimensions across countries.}
\begin{tabular}{|l|r|r|r|}
\hline
& \multicolumn{3}{|c|}{\textbf{chi-square test p-value}} \\ \cline{2-4} 
\multicolumn{1}{|c|}{\textbf{emotion}} & \multicolumn{1}{|c|}{\textbf{hue}} & \multicolumn{1}{|c|}{\textbf{saturation}} & \multicolumn{1}{|c|}{\textbf{value}}  \\
\hline
anger & 0.3942 & 0.4629 & 0.4792  \\ \hline
contempt & 0.5107 & 0.3811 & 0.2499  \\ \hline
disgust  &  0.4487 & 0.3987 & 0.3200  \\ \hline
fear & 0.4309 & 0.4870 & 0.4467  \\ \hline
guilt  & 0.3070 & 0.4198 & 0.2444  \\ \hline
interest & 0.3443 & 0.4479 & 0.4300  \\ \hline
joy & 0.3882 & 0.4713 & 0.5187  \\ \hline
sadness & 0.4251 & 0.4050 & 0.3262  \\ \hline
shyness & 0.3537 & 0.1060 & 0.3417 \\ \hline
surprise & 0.3896 & 0.4473 & 0.3853   \\ \hline
\end{tabular}
\label{tab:chi2_color}
\end{table}

\begin{table}[tb]
\centering
\caption{Independence of animation parameters across countries (* - significant difference).}
\begin{tabular}{|l|r|r|r|r|}
\hline
& \multicolumn{4}{|c|}{\textbf{chi-square test p-value}} \\ \cline{2-5}
\multicolumn{1}{|c|}{\textbf{emotion}}& \multicolumn{1}{|c|}{\textbf{speed}} & \multicolumn{1}{|c|}{\textbf{size}} & \multicolumn{1}{|c|}{\textbf{shape}} & \multicolumn{1}{|c|}{\textbf{animation}} \\
\hline
anger & 0.0830 & 0.1499 & 0.2534 & 0.1813  \\ \hline
contempt & * 0.0397 & 0.8051 & 0.1343 & 0.7418  \\ \hline
disgust & 0.3614 &0.1398 & 0.9992 & 0.6594  \\ \hline
fear & 0.1009 & 0.4718 & 0.1956 & 0.3080  \\ \hline
guilt & 0.1803 & 0.2157 & 0.6160 & 0.2395  \\ \hline
interest & 0.2402 & 0.8534 & 0.2582 & 0.3649  \\ \hline
joy & * 0.0442 & 0.2034 & 0.0768 & 0.2667  \\ \hline
sadness & * 0.0293 & 0.1454 & 0.4251 & 0.5329  \\ \hline
shyness & * 0.0002 & 0.2271 & * 0.0278 & * 0.0056  \\ \hline
surprise & 0.1404 & 0.2574 & 0.7889 & 0.2420  \\ \hline
\end{tabular}

\label{tab:chi2}
\end{table}

A nonparametric chi-square test was employed to assess the independence of variables within a contingency table across cultural groups, and the results are presented in Table~\ref{tab:chi2_color} and  Table~\ref{tab:chi2}. A significance threshold of 0.05 was set to reject the null hypothesis indicating a relationship between cultural background and the selected parameters.  Conversely, higher p-values indicate greater similarity in responses between the two groups, supporting the hypothesis that certain animation properties are culturally universal.

The analysis of cultural differences in the representation of emotions through animation parameters revealed varying degrees of universality. The results, shown in Table~\ref{tab:chi2_color}, suggest that color can serve as a universal medium for emotion representation, as no statistically significant differences were observed between Polish and Turkish participants regarding color components, i.e. hue, value and saturation. For all emotions examined, the p-values exceeded 0.05, with the lowest p-value observed for saturation in the context of the emotion of shyness. 

In addition to color-related parameters, size also showed cultural universality, as no statistically significant differences were observed between Polish and Turkish participants for any emotion (Table~~\ref{tab:chi2}). The lowest p-values for size were found for disgust (p = 0.1398), sadness (p = 0.1454), and anger (p = 0.1499); however, these values remain above the threshold of 0.05, indicating that object size in animated representations is consistent across both cultural groups.

In contrast to color and size, certain parameters for specific emotions showed notable cultural differences. Shape exhibited some degree of cultural divergence, particularly for the emotion of shyness (p=0.0278). This suggests that the association of certain geometric shapes with certain emotions may be influenced by cultural context. Furthermore, animation type revealed significant cross-cultural differences also for the emotion of shyness (p=0.0056). Although there is a significant difference for shape and animation type only for one emotion, the results question the universality of these animation parameters as they can be influenced by cultural factors.

Chi-squared analysis of the speed-dependent variable revealed the most pronounced significant differences between participants from the countries studied. For three of the emotions studied, the p-values were less than 0.05, indicating significant cultural differences in the perception of speed as an emotional cue. For the remaining emotions, the p-values were also relatively low, with the sole exception of disgust, which yielded a p-value greater than 0.25. To further investigate the source of these differences, the mean and standard deviation of speed ratings were calculated separately for each cultural group, as detailed in Table~\ref{tab:speed_countries}. 

\begin{table}[tb]
\centering
\caption{Descriptive statistics for indicated speed values for each emotion across countries.}
\begin{tabular}{|l|r|r|r|r|}
\hline
 & \multicolumn{2}{|c|}{\textbf{Turkey}} & \multicolumn{2}{|c|}{\textbf{Poland}} \\ \cline{2-5}

\textbf{emotion} & \multicolumn{1}{|c|}{\textbf{mean}} & \multicolumn{1}{|c|}{\textbf{std. dev.}} 
& \multicolumn{1}{|c|}{\textbf{mean}} & \multicolumn{1}{|c|}{\textbf{std. dev.}} \\
\hline
anger & 0.7278 & 0.2955 & 0.8922 & 0.1998 \\ \hline
contempt & 0.5774 & 0.2978 & 0.3961 & 0.2698 \\ \hline
disgust & 0.5604 & 0.3002 & 0.5078 & 0.2606 \\ \hline
fear & 0.6370 & 0.3073 & 0.7882 & 0.2951 \\ \hline
guilt & 0.4377 & 0.3014 & 0.3902 & 0.2982 \\ \hline
interest & 0.5962 & 0.2780 & 0.5118 & 0.2688 \\ \hline
joy & 0.7519 & 0.2353 & 0.7392 & 0.2871 \\ \hline
sadness & 0.4037 & 0.2920 & 0.2137 & 0.1649 \\ \hline
shyness & 0.4962 & 0.3069 & 0.2294 & 0.2369 \\ \hline
surprise & 0.6963 & 0.2990 & 0.7000 & 0.2585 \\ \hline
\end{tabular}
\label{tab:speed_countries}
\end{table}

The results show that Polish participants tended to choose more extreme speed values for excitement, while Turkish participants operated more in the range of average values. For instance, in the case of anger, an emotion strongly linked to dynamic motion, the mean speed value chosen by Polish participants was 0.8922, compared to 0.7278 among Turkish participants. At the other end of the spectrum for emotions characterized by lower arousal, such as sadness and shyness, Polish participants selected considerably lower speed values, with means of 0.2137 and 0.2294, respectively. In contrast, Turkish participants assigned higher values for these emotions, averaging 0.4037 for sadness and 0.4962 for shyness.

\subsection{Correlations}
In order to propose a universal representation of emotions, respondents' responses were analyzed to examine the correlations between specific emotions and animation parameters. Statistical measures, including mean, standard deviation, and confidence intervals, were used to assess the consistency of the results. This analytical approach facilitated the identification of robust and generalizable patterns, contributing to a standardized framework for representing emotions through animation parameters.

\subsubsection{Colors}
In contrast to previous studies in which participants selected colors from a predefined and limited set~\cite{Wrobel2024, Jonauskaite2020Sep, Lin2021May}, this study allowed participants to choose colors freely within the RGB model, providing a more nuanced and detailed representation of their emotional associations. This approach provided greater flexibility and precision in capturing individual perceptions of color-emotion relationships. Subsequently, converting these RGB color selections to the HSV (hue, saturation, value) model facilitated more in-depth analysis, allowing for a comprehensive examination of how specific color attributes correlated with emotions. This methodological advance not only increased the granularity of the data but also allowed for a robust comparison of participants' color choices.

\begin{table*}[tb]
\centering
\caption{Descriptive statistics for indicated HSV color dimensions for each emotion. }
\begin{tabular}{|l|r|r|c|r|r|c|r|r|c|}
\hline
 & \multicolumn{3}{|c|}{\textbf{hue}} & \multicolumn{3}{|c|}{\textbf{saturation}}  & \multicolumn{3}{|c|}{\textbf{value}}\\
 \cline{2-10}
\textbf{emotion} & \multicolumn{1}{|c|}{\textbf{mean}} & \multicolumn{1}{|c|}{\textbf{std. dev.}} 
& \multicolumn{1}{|c|}{\textbf{CI}} 
& \multicolumn{1}{|c|}{\textbf{mean}} & \multicolumn{1}{|c|}{\textbf{std. dev.}} 
& \multicolumn{1}{|c|}{\textbf{CI}} 
& \multicolumn{1}{|c|}{\textbf{mean}} & \multicolumn{1}{|c|}{\textbf{std. dev.}} 
& \multicolumn{1}{|c|}{\textbf{CI}} \\
\hline

anger & 359.023 & 18.114 & 355.535 -- 2.512 & 91.352 & 22.685 & 86.983 -- 95.721 & 83.452 & 26.078 & 78.429 -- 88.474 \\ \hline
contempt & 354.707 & 74.105 & 340.366 -- 9.048 & 81.319 & 28.958 & 75.715 -- 86.923 & 69.098 & 32.145 & 62.877 -- 75.319 \\ \hline
disgust & 76.362 & 67.546 & 63.290 -- 89.434 & 76.599 & 27.803 & 71.219 -- 81.980 & 62.155 & 29.012 & 56.540 -- 67.770 \\ \hline
fear & 345.104 & 67.942 & 332.019 -- 358.188 & 61.338 & 42.990 & 53.058 -- 69.617 & 52.027 & 39.342 & 44.451 -- 59.604 \\ \hline
guilt & 343.276 & 67.727 & 330.169 -- 356.383 & 68.719 & 38.258 & 61.315 -- 76.123 & 66.495 & 33.935 & 59.928 -- 73.062 \\ \hline
interest & 142.448 & 112.716 & 120.635 -- 164.262 & 88.598 & 18.549 & 85.008 -- 92.188 & 88.631 & 15.425 & 85.646 -- 91.616 \\ \hline
joy & 90.042 & 76.666 & 75.277 -- 104.807 & 93.061 & 14.125 & 90.341 -- 95.781 & 91.461 & 14.135 & 88.739 -- 94.184 \\ \hline
sadness & 254.787 & 79.370 & 239.501 -- 270.072 & 65.195 & 39.770 & 57.536 -- 72.854 & 67.869 & 34.957 & 61.137 -- 74.601 \\ \hline
shyness & 337.066 & 99.564 & 317.798 -- 356.334 & 65.333 & 38.652 & 57.853 -- 72.814 & 80.732 & 23.110 & 76.260 -- 85.204 \\ \hline
surprise & 38.246 & 91.868 & 20.554 -- 55.939 & 90.296 & 18.948 & 86.647 -- 93.945 & 92.682 & 13.070 & 90.165 -- 95.199 \\ \hline
\end{tabular}
\label{tab:colors_statistics}
\end{table*}

To analyze the relationship between colors and emotions, the hue, saturation, and value (HSV) components were examined using statistical measures, including mean, standard deviation (std.dev.), and 95\% confidence intervals (CI). The mean provided an estimate of central tendency, indicating the most frequently selected value for each component across participants. The standard deviation measured the variability of responses, indicating whether participants exhibited strong agreement or diverse choices for a given emotion. Confidence intervals were used to assess the reliability of the estimated means, ensuring that observed patterns were not due to random variation. This combination of statistical measures allowed for a comprehensive assessment of trends in color choices while accounting for both consistency and variability within the dataset.

Table~\ref{tab:colors_statistics} presents the statistical analysis of the HSV values for the colors selected to represent each emotion. The analysis of hue about emotions revealed clear and distinctive color associations across various emotional categories. Negative emotions, such as anger (with mean hue value of 359.023\textdegree), contempt (354.707\textdegree), fear (345.104\textdegree), guilt (343.276\textdegree), and shyness (337.066)\textdegree, were predominantly linked to shades of red. 

In contrast, positive emotions like joy (90.042\textdegree) and interest (142.448\textdegree) were connected to greenish hues. Sadness (254.787\textdegree) was predominantly associated with a darker blue hue. Surprise (38.246\textdegree) was represented by orange, a warm and vibrant hue. Lastly, disgust (76.362\textdegree) was linked to khaki, a muted, earthy tone.

The hue values for the emotion of interest exhibited the widest 95\% confidence interval (120.635\textdegree -- 164.262\textdegree) among all emotions studied, indicating greater variability in participants' color choices. Despite this variability, both the lower and upper bounds of the interval fall within the green spectrum, suggesting a consistent association of interest with greenish hues. The lower value (120.635\textdegree) corresponds to a more vibrant, saturated green, while the higher value (164.262\textdegree) represents a softer, more toned green. The consistency within the green spectrum, despite the variability, underscores the robustness of green as a color associated with interest, while the range highlights individual differences in emotional expression. For all other emotions, the confidence intervals for hue were notably tighter, indicating greater consistency in participants' color choices. 

Analyzing the participants' saturation indications also reveals a certain relationship. Emotions with high dominance, such as anger, interest, joy, or surprise, are characterized by high saturation, with an average response of 91.352\%, 88.598\%, 93.061\%, and 90.296\%, respectively. In contrast, the emotions associated with withdrawal or being under control had a much lower response value, i.e., 65.333\% for shyness, 65.195\% for sadness, 61.338\% for fear, and 68.719\% for guilt. 

Additionally, the analysis of confidence intervals indicates that participants' responses were relatively consistent across emotions. While the intervals for fear, guilt, shyness, and sadness were somewhat wider, reflecting slightly greater variability, these values do not undermine the overall assumption of response consistency. 

The last component of the HSV model, value, which represents brightness, was generally low for emotions associated with withdrawal or avoidance, such as disgust, fear, guilt, and sadness. This is consistent with their muted and negative nature, as darker tones often reflect low arousal states. However, the high value for shyness is an unexpected finding, as it contrasts with the typical pattern of low brightness for withdrawal-related emotions. This suggests that shyness may be perceived differently in terms of lightness, perhaps reflecting its unique blend of social inhibition and self-consciousness. The higher value may indicate that shyness is associated with a pale or light color rather than a dark one, emphasizing its nuanced emotional quality as a softer, less intense form of withdrawal.

\begin{figure*}[tbp]
\centering
\includegraphics[width=0.55\textwidth]{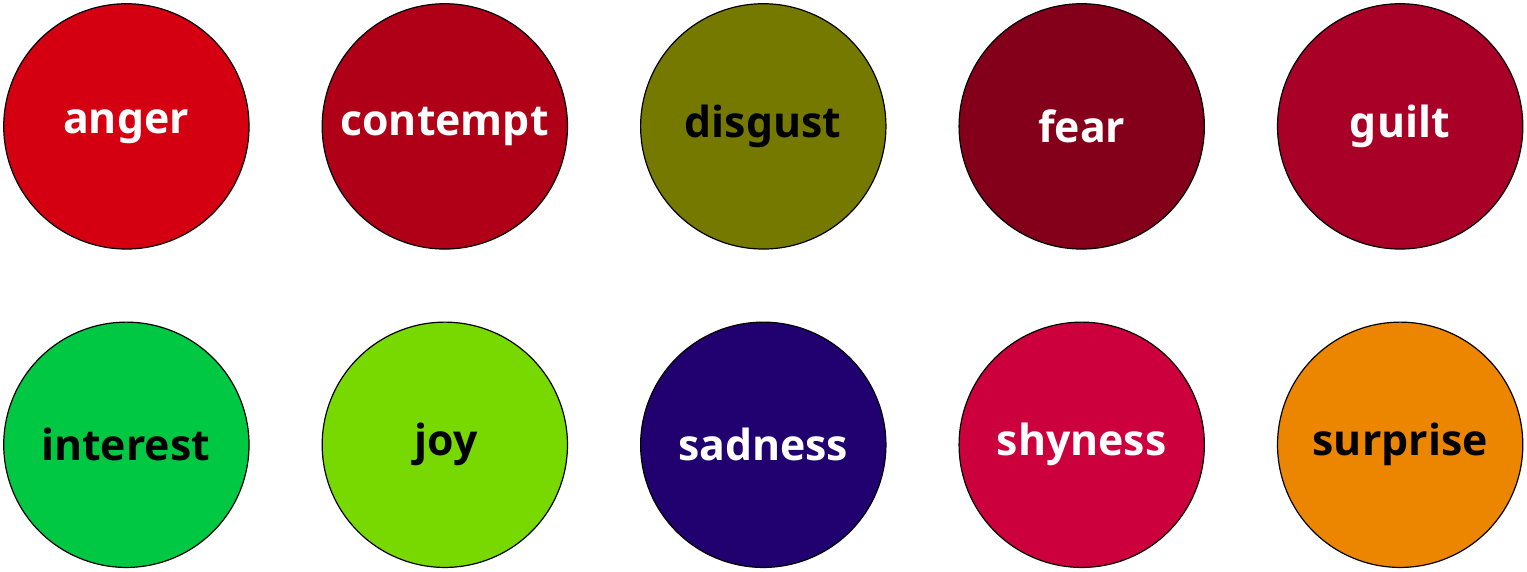}
\caption{Colors indications for each emotion.}
\label{fig:colors_result}
\end{figure*}

Figure~\ref{fig:colors_result} shows the colors generated for each of the studied emotions, derived by calculating the mean values of hue, saturation, and value (HSV) from participant responses. The mean hue determines the dominant tone, while saturation and value reflect the intensity and brightness of the selected colors. By averaging these values across participants, a representative color for each emotion was obtained, summarizing collective emotion-color associations.

The resulting colors reveal some common pattern in how emotions are visually represented by colors. High-arousal negative emotions such as anger, contempt, and fear are associated with deep red hues, indicating a strong link between warm colors and intense, aversive emotions. In contrast, joy and interest are represented by bright green shades, reinforcing their connection to positive affective states. Sadness and disgust appear in darker or muted tones, with sadness leaning toward deep blue and disgust toward green-brown, reflecting their lower arousal and negative valence. Shyness, depicted in lighter red-pink, may be perceived as a less intense or socially nuanced emotion. Surprise is represented by vivid orange, emphasizing its energetic and attention-grabbing nature. 

\subsubsection{Speed and size}

\begin{table}[tb]
\centering
\caption{Confidence interval for indicated speed and size values associated with each emotion. }
\begin{tabular}{|l|c|c|}
\hline
\textbf{emotion} & \textbf{size}& \textbf{speed} \\
\hline
anger & 0.739 -- 0.840 & 0.752 -- 0.855 \\ \hline
contempt & 0.558 -- 0.655& 0.429 -- 0.544 \\ \hline
disgust &  0.518 -- 0.606 & 0.479 -- 0.588 \\ \hline
fear & 0.446 -- 0.562 & 0.648 -- 0.767 \\ \hline
guilt & 0.462 -- 0.561& 0.357 -- 0.472 \\ \hline
interest & 0.560 -- 0.649 & 0.500 -- 0.606 \\ \hline
joy & 0.602 -- 0.702& 0.691 -- 0.792 \\ \hline
sadness & 0.454 -- 0.552& 0.263 -- 0.361 \\ \hline
shyness & 0.281 -- 0.376& 0.306 -- 0.423 \\ \hline
surprise & 0.590 -- 0.672& 0.646 -- 0.753 \\ \hline
\end{tabular}
\label{tab:size_speed_discrete}
\end{table}

The analysis of size and speed in relation to emotions revealed patterns consistent with the concepts of dominance and arousal. Table~\ref{tab:size_speed_discrete} presents confidence intervals showing that emotions associated with high dominance, such as anger (with mean 0.790), joy (0.652), and surprise (0.631), were consistently associated with larger sizes, reflecting their assertive and powerful nature. In contrast, emotions associated with low dominance, such as sadness (0.503), shyness (0.329), and fear (0.504), were associated with smaller sizes, consistent with their passive and submissive qualities. The narrow confidence intervals for most emotions indicate strong agreement among participants, reinforcing the relationship between size and perceived emotional dominance.

For speed, emotions characterized by high arousal, such as anger (0.804), fear (0.708), and joy (0.742), were associated with faster speeds, and reflecting their energetic and dynamic nature. In contrast, emotions associated with low arousal, such as sadness (0.312) and shyness (0.365), were associated with slower speeds, consistent with their restrained and withdrawn nature. Confidence intervals for speed were generally narrow for high arousal emotions, indicating strong consensus, while low arousal emotions had slightly wider intervals, suggesting greater variability in responses.

\subsubsection{Shape and animation type}

\begin{figure*}[tbp]
\centering
\includegraphics[width=0.93\textwidth]{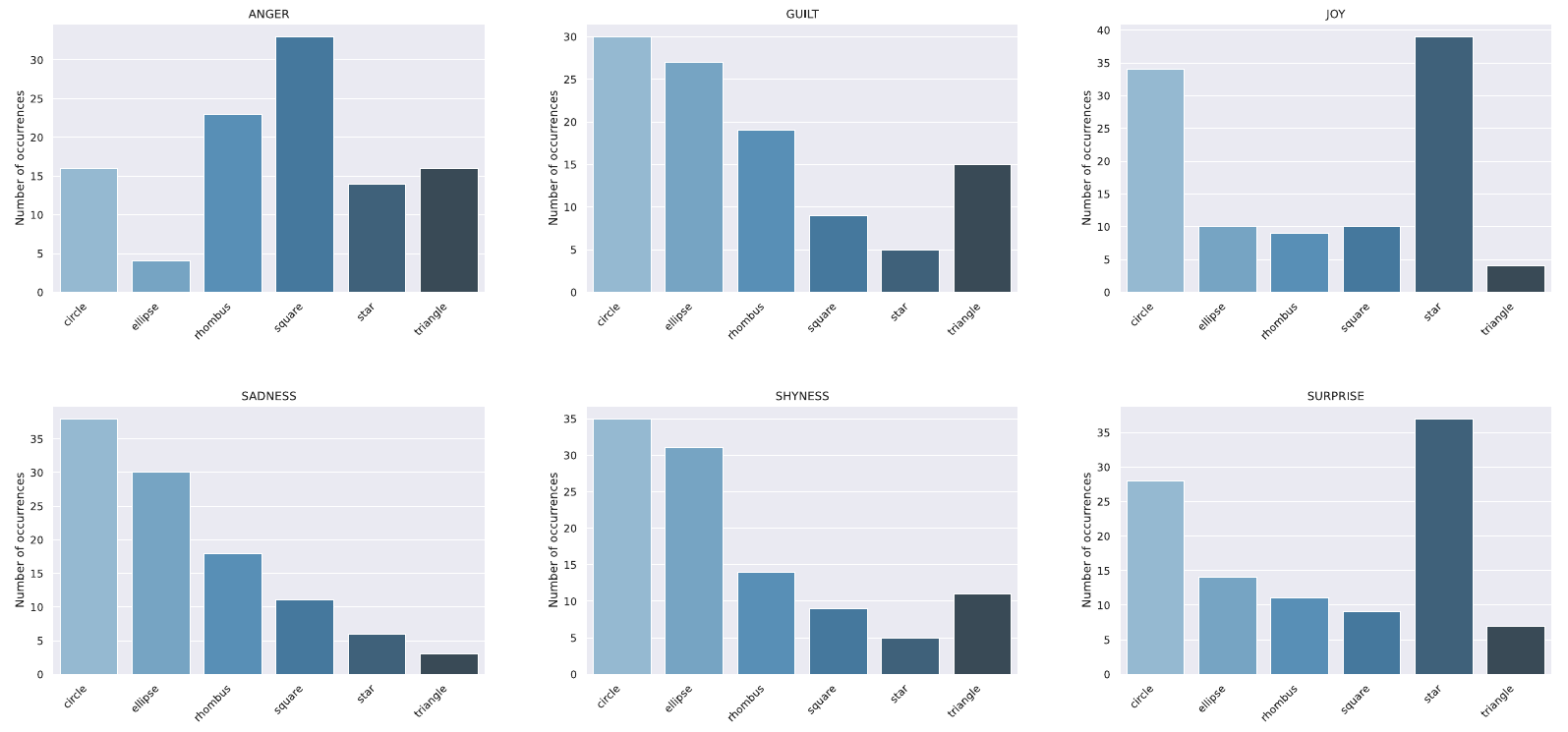}
\caption{Shape indications for selected six emotions.}
\label{fig:shape_result}
\end{figure*}

Figure~\ref{fig:shape_result} presents a bar chart illustrating the distribution of animation cues for the six emotions with the most distinct preferences, highlighting the clear associations between specific figure shapes and emotional states. 
High arousal emotions like joy and surprise are most commonly linked with the star shape, receiving 39 and 37 indications, respectively, followed closely by the circle (34 for joy and 28 for surprise). In contrast, low-arousal emotions like shyness and sadness show a preference for the circle and ellipse shapes, with the circle receiving 35 indications for shyness, and 38 for sadness, and the ellipse receiving 31 indications for shyness and 30 for sadness. On the other hand, anger, a strongly negative emotion, was closely associated with the square (33 indications), followed by the rhombus (23 indications), both sharp shapes. 

\begin{figure*}[tbp]
\centering
\includegraphics[width=0.93\textwidth]{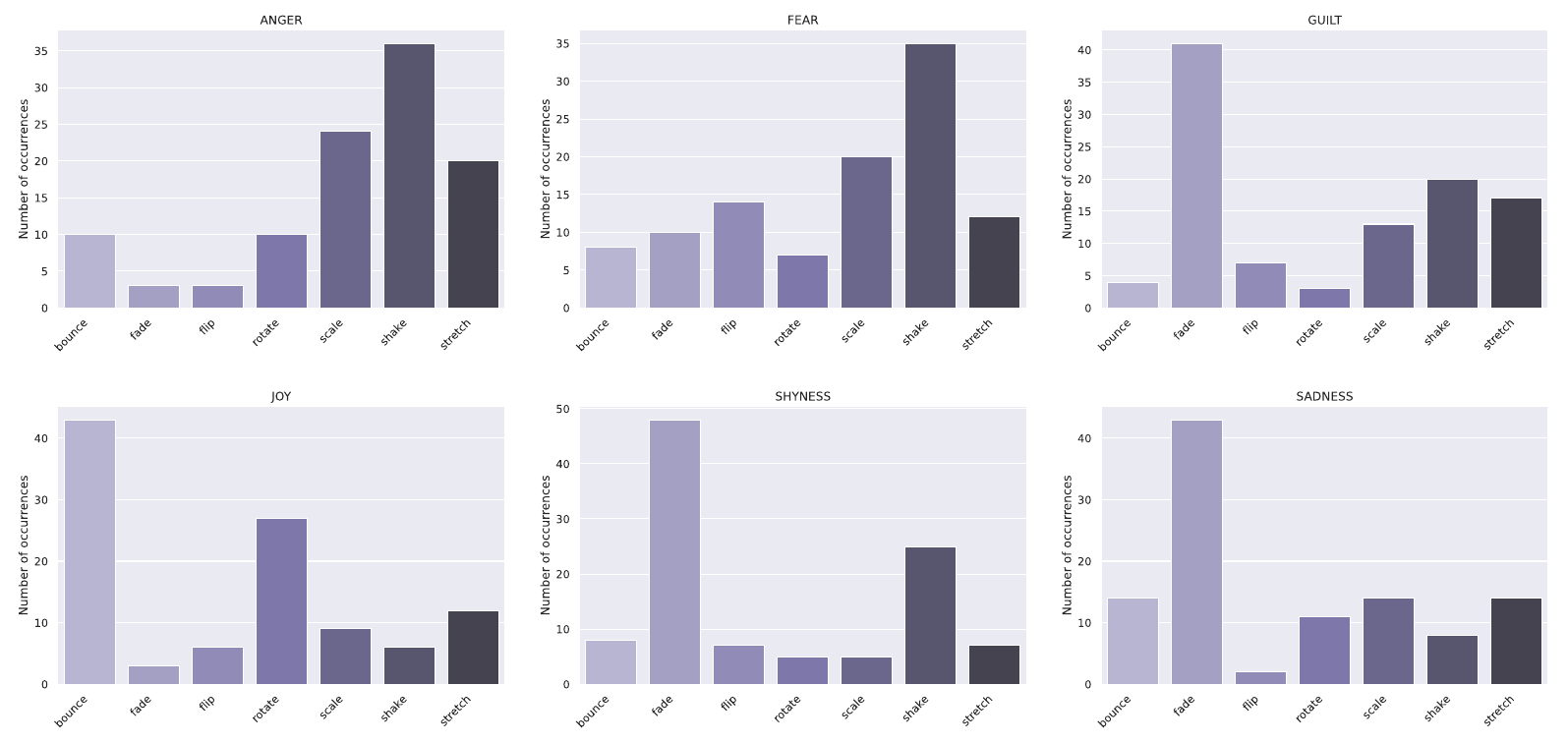}
\caption{Animation indications for selected six emotions.}
\label{fig:animation_result}
\end{figure*}

Analysis of animation type preferences across emotions revealed distinct patterns in how participants associated specific animations with emotional states, as shown in Figure~\ref{fig:animation_result}.
High arousal emotions such as anger and fear were strongly associated with dynamic and intense animations such as shake (36 and 35 indications, respectively) and scale (24 and 20 indications, respectively), reflecting their energetic and forceful nature. Similarly, joy was predominantly associated with bounce (43 indications), emphasizing its lively and exuberant character. In contrast, low-arousal emotions such as shyness, sadness, and guilt showed a preference for more subdued animations associated with fade, consistent with their passive and introspective qualities. Emotions such as interest and surprise showed a more balanced distribution across animations.

\section{Discussion}\label{sec:discussion}
The results provide valuable insights for designing pervasive affective computing systems that communicate sensed data back to users. By analyzing visual preferences among Polish and Turkish groups, we identified parameters that enable universal visualization versus visualization that requires cultural adaptation.

Color (hue, saturation, and value) and object size emerged as culturally independent variables, with no significant differences found between the two groups. For developers of wearable or ambient displays, these results suggest that color is a robust primary channel for encoding emotion. Red was predominantly associated with high-arousal negative emotions such as anger, contempt, and fear, while positive emotions such as joy and interest were associated with greenish hues. Sadness was most often represented by dark blue, consistent with its characterization as a low-arousal, negative emotion. In contrast, the surprise was associated with orange, a bright and attention-grabbing color. Furthermore, emotions associated with high dominance, such as anger and joy, were characterized by high color saturation, while emotions associated with withdrawal, such as sadness and fear, correlated with lower saturation and darker hues. Overall, universal algorithm can safely map positive valence (as detected by sensors) to greenish hues and negative valence to red or dark blue without risking misinterpretation. 

The results also confirmed that size may be linked to the emotional dimensions of dominance. Larger sizes were consistently associated with high-dominance emotions such as anger, joy, and surprise, underscoring their assertive and impactful characteristics. In contrast, smaller sizes were associated with low-dominance emotions, including sadness, shyness, and fear, reflecting their passive and subdued qualities. 

In contrast, animation speed revealed some cross-cultural differences. Although both groups associated faster speeds with higher arousal, the magnitude of this association differed. Polish participants used a wider dynamic range, preferring extreme speeds for high arousal. Turkish participants, on the other hand, preferred a more compressed, balanced range.
This finding has significant implications for biofeedback algorithms. A system that linearly maps physiological arousal (e.g., heart rate variability) to animation speed might appear ``hyperactive'' or jarring to a Turkish user but appropriate to a Polish user, or vice versa. Therefore, motion dynamics in affective interfaces cannot be purely universal. They require culturally adaptive calibration or user-specific baselines to ensure the feedback feels natural and unobtrusive.

\subsection{Limitations}
Participants in the study were primarily from Polish and Turkish cultural backgrounds, which may introduce sampling bias. While this allowed for a cross-cultural comparison, the findings may not apply to other, more distanced cultural groups, particularly those with different emotional expression norms or color associations. Additionally, the homogeneity of the sample, which exclusively included university students, and young and well-educated individuals, is the limitation of the study. This narrow demographic may not fully represent the broader population, as emotional perception and its visual representation could vary across different age groups, educational backgrounds, and cultural experiences. 

In this study, we used a predefined set of 10 emotion labels to ensure consistency with previous research. However, this selection may limit the generalizability of the findings because it does not cover the full range of human emotions, including more nuanced or culturally specific emotions.

\section{Conclusion}\label{sec:conclusion}
This study examined the potential of using abstract geometric animations as a universal language for expressing emotions on wearable devices with small screens. Through an analysis of user preferences among Polish and Turkish populations, we identified a robust set of visual parameters (color and size) that carry emotional meaning universally across cultures. These elements provide a stable foundation for designing global pervasive systems where specific sensor inputs, such as detected high arousal, can be reliably mapped to visual outputs, such as large, saturated shapes, without the risk of cross-cultural misinterpretation.

However, our findings also highlight a critical challenge for the algorithmic generation of feedback: the cultural relativity of speed. While the correlation between speed and arousal is universal, how this correlation is calibrated varies. Polish participants used a broader dynamic range for speed than Turkish participants, who had more balanced preferences. These results suggest that ``one-size-fits-all'' visualization algorithms may not resonate equally across populations. Thus, the development of affective computing systems requires the incorporation of cultural adaptation layers.

This work supports the development of generative visualization models that translate continuous physiological signals into intuitive, rapidly interpretable feedback. By mapping sensor-derived valence to color hue, arousal to saturation, and animation speed, pervasive devices can unobtrusively communicate complex emotional states. As we advance toward ubiquitous emotion sensing, it becomes clear that the final stage of the system -- the visual interface -- must be calibrated not only to the data, but also to the user's cultural context.

\bibliographystyle{IEEEtran}
\bibliography{bibliography.bib}


\end{document}